\documentclass[twocolumn,showpacs,aps,floatfix,superscriptaddress]{revtex4}
\usepackage{amsmath,amssymb,graphicx,eucal}
\begin{document}
\title{First Passage in Infinite Paraboloidal Domains}
\author{P. L. Krapivsky}
\affiliation{Center for Polymer Studies and Department of Physics, Boston University, Boston,
MA 02215, USA}
\affiliation{Institut de Physique Th\'eorique CEA, IPhT, F-91191 Gif-sur-Yvette, France}
\author{S. Redner}
\affiliation{Center for Polymer Studies and Department of Physics, Boston
  University, Boston, MA 02215, USA}

\begin{abstract}
  We study first-passage properties for a particle that diffuses either
  inside or outside of generalized paraboloids, defined by 
  $y=a(x_1^2+\ldots+x_{d-1}^2)^{p/2}$ where $p>1$, with
  absorbing boundaries.  When the particle is inside the paraboloid, the
  survival probability $S(t)$ generically decays as a stretched exponential,
  $\ln S\sim -t^{(p-1)/(p+1)}$, independent of the spatial dimensional.  For
  a particle outside the paraboloid, the dimensionality governs the
  asymptotic decay, while the exponent $p$ specifying the paraboloid is
  irrelevant.  In two and three dimensions, $S\sim t^{-1/4}$ and $S\sim (\ln
  t)^{-1}$, respectively, while in higher dimensions the particle survives
  with a finite probability.  We also investigate the situation where the
  interior of a paraboloid is uniformly filled with non-interacting diffusing
  particles and estimate the distance between the closest surviving particle
  and the apex of the paraboloid.
\end{abstract}
\pacs{02.50.Cw, 05.40.-a, 05.40.Jc, 02.30.Em}
\maketitle

\section{Introduction}

Random walks and diffusion are used to model numerous phenomena in physics,
chemistry, and biology \cite{wf,hcb,w,rg}.  In many applications, a diffusing
particle is confined to a certain domain and is absorbed if it hits the
boundary of this domain.  A basic problem is to determine the survival
probability of the particle~\cite{sr}.  For finite domains, the survival
probability decays exponentially with time, while richer behaviors may occur
for unbounded domains.  Among the possibilities for such infinite domains,
cones have been predominantly studied in the
physics~\cite{mef,dba,fg,kr,ck,bk1,bk2} and
mathematics~\cite{hn,bg,rdd,dz,bs,bd,djg,ntv} literatures, both because of
their simplicity and also their applications to the survival probabilities of
three or more mutually annihilating random walkers in one
dimension~\cite{mef,dba,fg,kr,ck,bk2}.

\begin{figure}
\begin{center}
\includegraphics[width=0.325\textwidth]{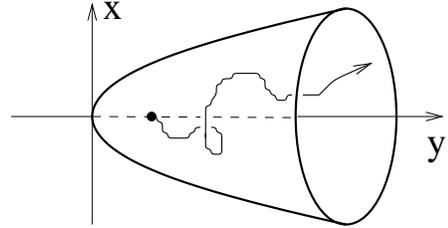}
\caption{A diffusing particle inside a paraboloid of revolution.  The
  particle starts along the symmetry axis and is absorbed when it hits the
  paraboloidal surface.}
\label{model}
\end{center}
\end{figure}

While a full understanding of diffusion inside cones is still incomplete, the
first-passage behavior of a diffusing particle inside a circular cone in any
dimension is well understood.  It is known that the survival probability
decays algebraically with time, and a good lower bound for the survival
exponent is also known \cite{bk1}.  In contrast, little is known about the
behavior of the survival probability in non-conical but still symmetric
infinite domains.  A natural appealing example of this latter class of
systems is that of a diffusing particle inside a paraboloid
(Fig.~\ref{model}).  In the case of a two-dimensional parabola, it was
recently found that the survival probability decays as a stretched
exponential in time~\cite{bds,ls}.  Moreover the exact amplitude in the
exponent of this stretched exponential decay was obtained exactly~\cite{ls}.

In this work, we present a simple extreme statistics argument (sometimes
called a Lifshitz tail argument) that allows one to `understand' this
behavior for the survival probability inside paraboloids.  Our goal is to
quantify first-passage phenomena for a diffusing particle that is initially
inside, as well as outside a paraboloid.  In addition, we determine the
temporal behavior of the closest surviving particle to the paraboloid apex
when its interior or exterior is uniformly filled with non-interacting
particles that are absorbed upon hitting the surface of the paraboloid.

The rest of this article is organized as follows. In Sec.~\ref{fp:inside}, we
outline an extreme statistics argument to determine the survival probability
of a diffusing particle inside a parabola and extend it to other infinite
domains with convex boundaries.  For domains that asymptotically cover an
infinitesimal fraction of space, e.g., for paraboloids, the survival
probability generically exhibits stretched exponential behavior.  If the
particle starts {\em outside} a paraboloid (Sec.~\ref{fp:outside}), it will
survive with a finite probability when the spatial dimensionality is $d\geq
4$, while in three dimensions the survival probability decays as $(\ln
t)^{-1}$.  We treat this latter problem using parabolic coordinates and a
quasi-static approach.

In Sec.~\ref{fp:time} we compute the average time for a diffusing particle to
hit the paraboloid when it starts inside. (For the particle outside a
paraboloid the average hitting time is infinite.)~ In Sec.~\ref{closest}, we
examine a related problem in which a paraboloidal domain is initially
uniformly filled with non-interacting diffusing particles that are absorbed
by the boundary.  Here we study the time dependence of the distance $\ell$
between the apex and the closest particle to the apex.  We show that this
distance is universal and scales as $t^{1/2}$ when the interior of the
paraboloid is filled by particles; if the particles are outside a
three-dimensional paraboloid, then $\ell\sim (\ln t)^{1/3}$.

\section{First Passage for diffusion inside a Paraboloid}
\label{fp:inside}

We start by considering a diffusing particle that starts at some point \cite{location}
inside the two-dimensional parabola
\begin{equation}
\label{parabola}
y=a x^2\,.
\end{equation}
We seek the survival probability $S(t)$ that this particle has not yet hit
the boundary of the parabola up to time $t$.  This survival probability was
shown to asymptotically decay as a stretched exponential in time~\cite{bds}
\begin{equation}
\label{St}
S\sim \exp\left[-A\,t^{1/3}\right]\,,
\end{equation}
with specified upper and lower bounds for the amplitude $A$.  This asymptotic
form \eqref{St} was recently proved and the amplitude $A$ was determined
explicitly~\cite{ls} (see below).

\begin{figure}
\begin{center}
\includegraphics[width=0.3\textwidth]{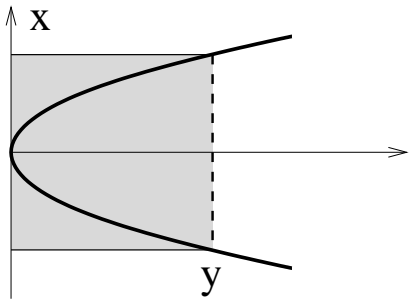}
\caption{\small The survival probability for a diffusing particle inside the
  parabola is approximated as the probability of remaining in the shaded
  rectangle and exiting along the dashed line.}
\label{bound}
\end{center}
\end{figure}

Let us try to understand the asymptotic \eqref{St} heuristically.  To this
end we construct an extreme statistics argument akin to that popularized in
the physics literature by its application to the random trapping
problem~\cite{BV,GP82} (see \cite{L64,KRB10} for reviews).  We make the
assertion that the asymptotic survival probability is controlled by the
probability that the particle wanders within the shaded rectangle of
Fig.~\ref{bound} and eventually exits this rectangle only along the dashed
boundary.  The probability for the particle to remain inside the interval
$(-x,x)$ scales as
\begin{equation*}
\exp\!\left[-\pi^2\,\frac{Dt}{(2x)^2}\right]\,.
\end{equation*}
Similarly, the probability for a particle to have longitudinal coordinate
$\geq y$, with $y\gg \sqrt{t}$, is governed by the factor
\begin{equation*}
\exp\!\left[-\frac{y^2}{4Dt}\right]\,.
\end{equation*}
For a particle with longitudinal coordinate $\geq y$, it will survive if its
transverse coordinate satisfies $|x|<\sqrt{y/a}$.  Combining the two factors
written above and writing $x= \sqrt{y/a}$, we may write the survival
probability as
\begin{equation}
\label{St:estimate}
S\alt\int_0^\infty dy\,\exp\!\left[-\frac{y^2}{4Dt} - \pi^2\,\frac{aDt}{4y}\right]\,.
\end{equation}
This expression represents an (asymptotic) upper bound to the true survival
probability because the particle may remain inside the rectangle but still
leave the parabola.

We now estimate the integral by the Laplace method by finding the maximum of
the integrand.  This maximum occurs at $y=y_*$ which is determined from the
condition
\begin{equation*}
2\,\frac{y_*}{4Dt} = \pi^2\,\frac{aDt}{4y_*^2}\,.
\end{equation*}
Therefore 
\begin{equation}
\label{St:est}
S\alt \exp\!\left[-\frac{3}{4}\,\left(\frac{a\pi^2}{2}\right)^{2/3} (Dt)^{1/3}\right]\,.
\end{equation}
This expression has the correct dependence on $t,D,a$, but the numerical
prefactor is incorrect.  According to~\cite{ls}, the amplitude is
$A_{\text{exact}}=\tfrac{3}{8}\pi^2$ when $a=1, D=\tfrac{1}{2}$, while
\eqref{St:est} gives $A=\tfrac{3}{8} \pi^{4/3}$.  Since $A<A_{\rm exact}$,
this argument gives an upper bound for the survival probability, consistent
with the approximation underlying our argument.

It is straightforward to extend the reasoning above to non-quadratic parabolas
that are defined by $y=a |x|^p$, with $p>1$.  The counterpart to
\eqref{St:estimate} is now
\begin{equation*}
S\sim \int dy\,\exp\!\left[-\frac{y^2}{4Dt} - \frac{\pi^2 Dt}{4}\,\left(\frac{a}{y}\right)^{2/p}\right]\,,
\end{equation*}
from which
\begin{equation}
\label{St:est-p}
S\sim \exp\!\left[-\frac{p+1}{4}\,\left(\frac{\pi^2 a^{2/p}}{p}\right)^{\frac{p}{p+1}} 
(Dt)^{\frac{p-1}{p+1}}\right]\,.
\end{equation}

The extreme statistics approach also works for other infinite convex
domains.  For example, if
\begin{equation}
\label{LL:exp}
y=L e^{(|x|/\ell)^b}\,,
\end{equation}
then the counterpart of Eq.~\eqref{St:estimate} is
\begin{equation*}
S\sim \int dy\,\exp\!\left[-\frac{y^2}{4Dt} - \frac{\pi^2 Dt}{4\ell^2}\,\left(\ln\frac{y}{L}\right)^{-2/b}\right]\,,
\end{equation*}
from which we obtain
\begin{equation*}
S\sim \exp\!\left[ - \frac{\pi^2 Dt}{4\ell^2}\,\left(\ln\frac{Dt}{L\ell}\right)^{-2/b}\right]\,.
\end{equation*}

A natural extension is to higher dimensions.  For example, for the
non-quadratic paraboloid in $d$ dimensions defined by
\begin{equation}
\label{p:paraboloid}
y=a R^p, \quad R=\sqrt{x_1^2+\ldots+x_{d-1}^2}, \quad p>1\,,
\end{equation}
we now obtain
\begin{equation}
S\sim \int_0^\infty dy\,\exp\!\left[-\frac{y^2}{4Dt} - \frac{j_{\delta}^2 Dt}{R^2}\right]\,.
\end{equation}
Here $j_\delta$ is the first positive zero of the Bessel function $J_\delta$,
with $\delta=\tfrac{d-3}{2}$.  This numerical factor arises from the
long-time solution to the diffusion equation inside a $(d-1)$-dimensional
sphere of radius $a$ with absorbing boundaries.  Finally, we substitute
$R=(y/a)^{1/p}$ into the above estimate and compute the dominant contribution
to the integral by the Laplace method to obtain
\begin{equation}
\label{St:est-pd}
S\sim \exp\!\left[-\frac{p+1}{4}\,\left(\frac{4j_{\delta}^2\,a^{2/p}}{p}\right)^{\frac{p}{p+1}} (Dt)^{\frac{p-1}{p+1}}\right]\,.
\end{equation}
Once again, the heuristic extreme statistics argument reproduces the correct
value of the exponent $\sigma=\frac{p-1}{p+1}$ in the stretched exponential~\cite{bds,ls},
but not the correct amplitude.

\section{First Passage for diffusion outside a Paraboloid}
\label{fp:outside}

Let us now study the complementary situation where a diffusing particle
starts {\em outside} a paraboloid.  This problem is reminiscent of diffusion
in (almost) free space except for the presence of an excluded half-line.
This analogy suggests that there should be a fundamental difference between
two and higher dimensions.  In two dimensions, the survival probability of a
diffusing particle in the presence of a semi-infinite absorbing line decays
as $t^{-1/4}$~\cite{bk1,CR89}.  The same time dependence applies for a
diffusing particle exterior to a parabola because the width of the parabola
is asymptotically negligible in comparison with the longitudinal coordinate.
On the other hand, when $d\geq 4$, a diffusing particle survives with
positive probability in the presence of an excluded semi-infinite ray that
has a non-zero width.  The same behavior should also occur when the excluded
region is a paraboloid.

The most interesting behavior occurs in three dimensions.  We shall see that
the survival probability of a diffusing particle in the exterior of the
three-dimensional paraboloid decays as
\begin{equation}
\label{St:outside}
S\sim [\ln(a^2Dt)]^{-1}\,,
\end{equation}
Similarly, for generalized three-dimensional paraboloids defined by
Eq.~\eqref{p:paraboloid}, the survival probability has essentially the
same form:
\begin{equation}
\label{St:outside-p}
S\sim [\ln(a^{2/(p-1)} Dt)]^{-1}\,.
\end{equation}

To establish \eqref{St:outside} and \eqref{St:outside-p}, we recall that the
survival probability satisfies the diffusion equation~\cite{sr}
\begin{equation*}
\frac{\partial S}{\partial t} = D \nabla^2 S, \quad S|_{\text{boundary}}=0, \quad S|_{t=0}=1\,.
\end{equation*}

To solve this boundary-value problem we employ the powerful yet simple
quasi-static approximation~\cite{RPJ77,C87,RA90,PLK93}.  The quasi-static
method is based on dropping the time derivative in the diffusion equation and
solving the resulting Laplace equation for distances $\leq \sqrt{Dt}$ from
the paraboloid and then matching this density to its unperturbed value when
the distance equals $\sqrt{Dt}$.  This approach is especially useful for
simple geometries, such as planar or cylindrical absorbing boundaries.

In the case of the paraboloid, the quasi-static approach requires a bit more
effort, as the boundary is not elementary.  However, the boundary simplifies
in parabolic coordinates, and the convenience of a simple boundary condition
offsets the complication of dealing with the Laplace equation in parabolic
coordinates.  The parabolic coordinates $(\xi,\eta,\phi)$ are related to the
Cartesian coordinates through
\begin{equation}
\label{parabolic_coor}
\begin{split}
x_1 &= \sqrt{\xi\eta}\, \cos\phi\,,\\
x_2 &= \sqrt{\xi\eta}\, \sin\phi\,,\\
y &= \tfrac{1}{2}(\xi - \eta)\,.
\end{split}
\end{equation}
with $\xi\geq 0, \eta\geq 0$ and $0\leq \phi\leq 2\pi$.  Here the families of
surfaces $\xi=\text{const}$ and $\eta=\text{const}$ are confocal paraboloids
whose common focus is the origin.

In parabolic coordinates, the Laplacian is
\begin{equation*}
\nabla^2 = \frac{4}{\xi+\eta}\left[\frac{\partial}{\partial \xi}\left(\xi\,\frac{\partial}{\partial \xi}\right)
+ \frac{\partial}{\partial \eta}\left(\eta\,\frac{\partial}{\partial \eta}\right)\right]+
\frac{1}{\xi \eta}\,\frac{\partial^2}{\partial \phi^2} \,,
\end{equation*}
and we wish to solve the Laplace equation $\nabla^2 S = 0$ in this coordinate
system.  The solution has axial symmetry $S = S(\xi,\eta)$.  By defining the
paraboloid in the form $\xi=\xi_0$, the boundary condition $S(\xi_0,\eta)=0$
suggests seeking a solution as a function of $\xi$ alone, $S = S(\xi)$.  For
this choice the Laplace equation $\nabla^2 S = 0$ reduces to
\begin{equation*}
\frac{d}{d \xi}\left(\xi\,\frac{d S}{d \xi}\right) = 0\,,
\end{equation*}
whose solution is $S = C \ln(\xi/\xi_0)$.  Matching this solution to the
initial density at $\xi\sim \sqrt{Dt}$ fixes the constant and the full
solution is
\begin{equation}
\label{St:paraboloid}
S = \frac{\ln(\xi/\xi_0)}{\ln(\sqrt{Dt}/\xi_0)}\,.
\end{equation}
(As in other applications of the quasi-static approach, the `constant' $C$ is
actually time dependent.)~ If the diffusing particle starts not far from the
apex of the parabola, $\xi\sim \xi_0$, the survival probability
\eqref{St:paraboloid} becomes $S\sim [\ln(Dt/\xi_0^2)]^{-1}$, thereby
establishing~\eqref{St:outside}.  

Alternatively, Eq.~\eqref{St:outside} may be inferred from the well-known
survival probability of a diffusing particle exterior to an absorbing circle
in two dimensions~\cite{w,sr}.  As long as the diffusing particle remains in
the half-space $y>0$, the absorbing boundary is a circle that represents the
two-dimensional projection of the absorbing parabola.  The radius of the
circle is not fixed, but rather varies in time as $R\sim\sqrt{y}\sim
t^{1/4}$.  However, this slower than diffusive variation in the radius does
not affect the asymptotics of the survival probability.  This same approach
applies to the non-quadratic parabola $y=a R^p$.  The only new feature is
that the amplitude $a$ appears as $a^{2/(p-1)}$; this dependence is mandated
by dimensional analysis.  In this way we also establish
Eq.~\eqref{St:outside-p}.

When $d>3$, we use the $d-$dimensional generalization of parabolic
coordinates $(\xi,\eta,\phi_1,\ldots,\phi_{d-2})$ in which the angular
coordinates $(\phi_1,\ldots,\phi_{d-2})$ are the same as those for spherical
coordinates in $d-1$ spatial dimensions.  For example, in four dimensions
\begin{equation*}
\begin{split}
x_1 &= \sqrt{\xi\eta}\, \cos\phi_1\,,\\
x_2 &= \sqrt{\xi\eta}\, \sin\phi_1\cos\phi_2\,,\\
x_3 &= \sqrt{\xi\eta}\, \sin\phi_1\sin\phi_2\,,\\
y &= \tfrac{1}{2}(\xi - \eta)\,.
\end{split}
\end{equation*}
where again $\xi\geq 0, \eta\geq 0$ and the angular coordinates vary in the
range $0\leq \phi_1\leq \pi$ and $0\leq \phi_2\leq 2\pi$.

The Laplacian in $d$-dimensional parabolic coordinates is
\begin{equation*}
\nabla^2 \!=\! \frac{4}{\xi\!+\!\eta}\left[
\frac{1}{\xi^\delta}\,\frac{\partial}{\partial \xi}
\left(\!\xi^{1+\delta}\frac{\partial}{\partial \xi}\!\right)
+ \frac{1}{\eta^\delta}\,\frac{\partial}{\partial \eta}
\left(\!\eta^{1+\delta}\frac{\partial}{\partial \eta}\!\right)\right]+
\frac{\mathcal{L}}{\xi \eta}\,,
\end{equation*}
where $\mathcal{L}$ is the angular part of the Laplacian in $d-1$ spatial
dimensions and we again use the notation $\delta=\frac{d-3}{2}$.  Since the
problem is axisymmetric, the solution is independent of the angular
coordinates and additionally the survival probability depends only on $\xi$.
In this case, the Laplace equation $\nabla^2 S = 0$ reduces to
\begin{equation*}
\frac{d}{d \xi}\left(\xi^{1+\delta}\frac{d S}{d \xi}\right) = 0\,,
\end{equation*}
whose solution is
\begin{equation}
\label{S:finite}
S = 1-\left(\frac{\xi_0}{\xi}\right)^\delta\,.
\end{equation}
Therefore instead of decaying to zero, the survival probability remains finite.

Summarizing, we have 
\begin{equation}
\label{Std:out}
S \sim 
\begin{cases}
t^{-1/4}      & d=2\,,\\
(\ln t)^{-1}  & d=3\,,\\
\text{finite} & d\geq 4\,.
\end{cases}
\end{equation}
The qualitative behavior \eqref{Std:out} clearly continues to hold for the
generalized paraboloids \eqref{p:paraboloid}.

\section{First Passage Time}
\label{fp:time}

For a finite domain, and for various infinite domains such as paraboloids, a
diffusing particle is certain to reach the boundary, and its average hitting
time is finite.  For a diffusing particle inside such a domain, the hitting
time $t=t(x,y)$ is a random variable; here $(x,y)$ denotes the starting
position of the particle.  The hitting time $T(x,y)=\langle t(x,y)\rangle$
averaged over all trajectories that start from $(x,y)$ satisfies the Poisson
equation~\cite{sr,KRB10}
\begin{equation}
\label{T:poisson}
D \nabla^2 T = -1, \quad T|_{\text{boundary}}=0\,.
\end{equation}

Let us first determine the exit time for a diffusing particle inside the
parabola $y=ax^2$.  Setting $a=1$ we must solve $D\nabla^2 T = -1$ subject to
the boundary condition $T=0$ when $y=x^2$.  It is natural to choose
$C(y-x^2)$ as a trial solution, as this function automatically obeys the
adsorbing boundary condition and is positive (when $C>0$) inside the
parabola.  Substituting this trial function into \eqref{T:poisson} we find
that it represents a solution when $C=(2D)^{-1}$. Therefore
\begin{equation}
\label{T:parabola}
D\,T=\frac{1}{2}(y-x^2)\,.
\end{equation}
This Laplacian formalism can be straightforwardly extended to determine the
higher moments of the hitting time.  For example, to determine the second
moment $T_2=\langle t(x,y)^2\rangle$ we must solve the boundary-value
problem~\cite{sr}
\begin{equation}
\label{T2}
D\nabla^2 T_2 = -2T; \qquad T_2=0\quad \text{when}\quad y=x^2
\end{equation}
The form of the solution \eqref{T:parabola} for the first moment suggests
trying a polynomial solution that is divisible by $y-x^2$.  By trial and
error, the appropriate form of $T_2$ is
\begin{equation*}
T_2 = A(y-x^2)^2+By(y-x^2)+C(y-x^2)\,.
\end{equation*}
Substituting this ansatz into \eqref{T2} fixes the constants $A,B,C$ and we
obtain
\begin{equation}
\label{T2-sol}
D T_2 = \tfrac{1}{12 }(y-x^2)^2+\tfrac{1}{3}y(y-x^2)+\tfrac{5}{12}(y-x^2)\,.
\end{equation}
As might be expected, the second moment of the hitting time scales as the
square of the first moment.

For the $d-$dimensional paraboloid defined by the equation $y = R^2 \equiv
x_1^2+\ldots+x_{d-1}^2$, the corresponding results for the average hitting
time and the second moment are
\begin{align}
\begin{split}
T &= \frac{y - R^2}{2(d-1) D}\,,\\
T_2 &= \frac{y-R^2}{D(d^2\!-\!1)}\left[\left(\frac{3}{4}\!+\!\frac{1}{d}\right)y
\!-\!\frac{1}{4}\,R^2+\frac{3d\!+\!4}{2d^2}\right]\,.
\end{split}
\end{align}

\section{The closest particle}
\label{closest}

\begin{figure}[ht]
\begin{center}
\includegraphics[width=0.3\textwidth]{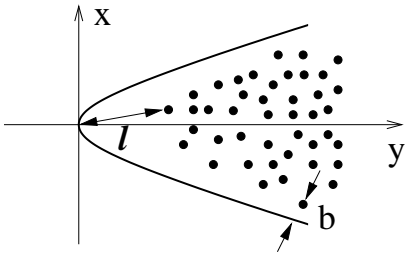}
\caption{Definition of the closest distances $\ell$ and $b$ for an initially
  constant density of diffusing particles inside an absorbing parabola.}
\label{l}
\end{center}
\end{figure}

Suppose that an infinite absorbing paraboloidal domain initially contains a
constant density $\rho$ of non-interacting diffusing particles.  As the
particles diffuse, the density near the boundary of the paraboloid is
depleted.  For particles that have not yet been absorbed, a natural way to
characterize their spatial distribution is by the distance between the
paraboloid and the closest particle.  For the paraboloidal geometry, there
are two natural (and distinct) definitions of closest particle
(Fig.~\ref{l}):
\begin{itemize}
\item the closest particle to the apex of the paraboloid, with corresponding
  distance $\ell$;
\item far from the apex, the closest particle to the side of the paraboloid
  (distance $b$ in Fig.~\ref{l}) is a natural measure of the depletion of the
  density.
\end{itemize}

To provide context for these quantities, let us recall basic features of the
corresponding problem for the simplest geometry of a constant density of
diffusing particles, with an absorbing spherical trap of finite radius $a$ at
the origin.  For this system, the distance $\ell$ to the closest diffusing
particle has the following time dependences~\cite{W89,RA90}
\begin{equation}
\ell\sim
\begin{cases} 
(Dt/\rho^2)^{1/4}&d=1 \,,\\
\sqrt{\ln(Dt/a^2)/(2\pi \rho)} & d=2\,,\\
\textrm{const.} & d>2\,.
\end{cases}
\end{equation}
Strikingly, the interaction between diffusion and absorption leads to a new
length scale $\ell$ that has a non-trivial time dependence and different
behavior as function of the spatial dimension.  Here we compute the
corresponding properties for diffusing particles inside a paraboloidal
absorbing boundary.

\subsection{Cones}

For circular cones, the asymptotics of the density distribution are known, so
that the closest particle problem admits an analytical solution.

\subsubsection{Wedge}

In two dimensions, the circular cone reduces to the wedge.  Let $\alpha$ be
the angle between the axis of the wedge and its surface.  Part of the reason
for first studying this system is that the limits of a narrow and a wide
wedge provide the asymptotics for the distance between the closest particle
and the apex of the parabola, both when the particle is inside or outside the
parabola.

Asymptotically, the density of surviving particles is given
by~\cite{bk1,CJ59}
\begin{equation}
\label{density}
\rho\left(\frac{r^2}{Dt}\right)^{\beta}\psi(\theta)\,,
\end{equation}
where $(r,\theta)$ are polar coordinates with the origin at the apex and with
$\theta=0$ corresponding to the symmetry axis of the wedge.  The angular
dependence of the density is $\psi(\theta)=\cos \tfrac{\pi\theta}{2\alpha}$
and the survival probability decays as $S\sim t^{-\beta}$, with exponent
$\beta=\tfrac{\pi}{4\alpha}$.

To estimate the distance $\ell\/$ from the apex of the wedge to the closest
surviving particle we use the extreme statistics criterion~\cite{KRB10}
\begin{equation}
\label{crit:2d}
\int_0^\ell r\,dr\int_{-\alpha}^\alpha d\theta\, \rho\left(\frac{r^2}{Dt}\right)^{\beta}\psi(\theta) \sim 1\,.
\end{equation}
Namely, there should be a single particle inside a pie-shaped sector of
length $\ell$ whose center is at the wedge apex.  Computing the integral we
obtain
\begin{align}
%\begin{split}
\label{dist:2d}
\ell%&\sim (Dt)^{\frac{\beta}{2\beta+2}}\, \rho^{-\frac{1}{2\beta+2}},, \\
\sim (Dt)^{\pi/(2\pi+8\alpha)}\,\rho^{-2\alpha/(\pi+4\alpha)}\,.
%\end{split}
\end{align}
Some interesting special cases are:
\begin{equation}
\label{close:2d}
\ell \sim 
\begin{cases}
(Dt)^{1/2}                        &\alpha\downarrow 0\,,\\
(Dt)^{1/4}\, \rho^{-1/4}    & \alpha=\pi/4\,,\\
(Dt)^{1/6} \,\rho^{-1/3}    & \alpha=\pi/2\,,\\
%(Dt)^{1/8}\, \rho^{-3/8}    & \alpha=3\pi/4\,,\\
(Dt)^{1/10}\, \rho^{-2/5}  & \alpha=\pi\,.
\end{cases}
\end{equation}
As the wedge becomes more open, it presents less of a ``hazard'' to diffusing
particles near the apex.  Thus the time dependence of $\ell$ becomes
progressively slower as $\alpha \to \pi$.

\subsubsection{Circular cone in $d$ dimensions}

For the $d$-dimensional circular cone, the density is still given by
Eq.~\eqref{density}, but the angular part of the density is now an associated
Legendre function (see, e.g., \cite{bk1}). The criterion \eqref{crit:2d}
generalizes to
\begin{equation}
\label{crit:d}
\int_0^\ell r^{d-1}\,dr\int_0^\alpha (\sin\theta)^{d-2}\,d\theta\, 
\rho\left(\frac{r^2}{Dt}\right)^{\beta}\psi(\theta) \sim 1\,,
\end{equation}
from which
\begin{equation}
\label{dist}
\ell\sim (Dt)^{\frac{\beta}{2\beta+d}}\, \rho^{-\frac{1}{2\beta+d}} \,.
\end{equation}
Here, the exponent $\beta$ and the opening angle of the cone $\alpha$ are
related by \cite{bk1}
\begin{equation}
\label{betad}
\begin{split}
P_{2\beta+\delta}^\delta(\cos\alpha) &= 0\qquad d\ {\rm odd}\,,\\
Q_{2\beta+\delta}^\delta(\cos\alpha) &= 0\qquad d\ {\rm even}\,.
\end{split}
\end{equation}
where $P_{2\beta+\delta}^\delta(\mu)$ and $Q_{2\beta+\delta}^\delta(\mu)$ are
the associated Legendre functions of degree $2\beta+\delta$ and order
$\delta$. Thus in general dimensions, the exponent $\beta$ is the root of a
transcendental equation \eqref{betad} that involves the associated Legendre
functions.  We must choose the smallest such root to ensure $\psi(\theta)>0$
for all $\theta<\alpha$.

In addition to two dimensions (the wedge), simple results arise in four
dimensions where 
\begin{equation*}
\psi_4(\theta)= \frac{\sin\left[(2\beta+1)\theta\right]}{\sin\theta}\,,
\end{equation*}
and the boundary condition $\psi(\alpha)=0$ gives  \cite{bk1}
\begin{equation}
\label{beta4}
\beta_4(\alpha)=\frac{\pi-\alpha}{2\alpha}\,.
\end{equation}
In four dimensions, Eq.~\eqref{dist} thus reduces to
\begin{equation}
\label{dist:4d}
\ell\sim (Dt)^{(\pi-\alpha)/(2\pi+6\alpha)}\, \rho^{-\alpha/(\pi+3\alpha)}\,.
\end{equation}
Some interesting special cases are:
\begin{equation}
\ell \sim 
\begin{cases}
(Dt)^{1/2}                            &\alpha\downarrow 0\,,\\
(Dt)^{3/14}\, \rho^{-1/7}      & \alpha=\pi/4\,,\\
(Dt)^{1/10}\,\rho^{-1/5}       & \alpha=\pi/2\,,\\
%(Dt)^{1/26}\,\rho^{-3/13}     & \alpha=3\pi/4\,,\\
\rho^{-1/4}                          & \alpha=\pi\,.
\end{cases}
\end{equation}

For general spatial dimension, the corresponding results are:
\begin{equation}
\label{l-general}
\ell \sim 
\begin{cases}
(Dt)^{1/2}                            &\alpha\downarrow 0\,,\\
(Dt/\rho)^{1/(2+d)}              & \alpha=\cos^{-1}(1/\sqrt{d})\,,\\
(Dt/\rho^2)^{1/(2+2d)}        & \alpha=\pi/2\,,\\
\rho^{-1/d}                          & \alpha=\pi ~~(d\geq 3)\,.
\end{cases}
\end{equation}
Each of these cases has an interesting interpretation.  For a very narrow
cone ($\alpha\downarrow 0$), the distance to the closest particle grows
diffusively, as $t^{1/2}$.  However, density dependence (Eqs.~\eqref{dist}
and \eqref{dist:4d}) leads to a divergent prefactor as $\alpha\downarrow 0$.
The case $\alpha=\cos^{-1}(1/\sqrt{d})$ corresponds to survival exponent
$\beta=1$~\cite{bk1}, that separates the regimes where the average survival
time is either finite (for $\alpha<\cos^{-1}(1/\sqrt{d})$ or divergent.  The
case $\alpha=\pi/2$ corresponds to an absorbing plane, and the behavior of
$\ell$ can be simply inferred from the criterion that there should be a
single particle within a hemispherical domain of radius $\ell$ about a point
on the boundary.  Since the density profile is a linear function of distance
to the boundary, this calculation is trivial.  Finally, when the cone becomes
the half-line ($\alpha=\pi$), there is a non-zero survival probability and
the distance between the closest particle and the apex merely reduces to the
average distance between particles.

\subsection{Paraboloids}

Let us now analyze the distance between the closest surviving diffusing
particle and the origin of a $d$-dimensional paraboloid.

\subsubsection{Particles inside the paraboloid}

Using the results for the circular cones, and taking the $\alpha\to 0$ limit,
gives the universal growth law
\begin{equation}
  \ell \sim \sqrt{Dt}\,.
\end{equation}
This prediction is valid in arbitrary dimension for all paraboloids,
both quadratic and higher order, as in \eqref{p:paraboloid}.

\subsubsection{Particles outside the paraboloid}

For the complementary situation where particles are uniformly distributed
outside an absorbing paraboloid, it is natural to adopt the results for the
circular cones in the $\alpha\to\pi$ limit.  This gives
\begin{equation}
\label{dist:naive}
\ell \sim 
\begin{cases}
(Dt)^{1/10}\, \rho^{-2/5}          &d=2\,,\\
\rho^{-1/d}                       &d\geq 3\,.
\end{cases}
\end{equation}
In two dimensions, $\ell$ slowly increases with time, a result that continues
to hold even for generalized parabolas \eqref{p:paraboloid}.  In three
dimensions, however, the behavior is more subtle because of the ultra-slow
logarithmic decay of the survival probability with time (see
Eqs.~\eqref{St:outside} and \eqref{St:outside-p}).  To investigate this case,
we use the quasi-static result \eqref{St:paraboloid} for the density,
\begin{equation*}
\rho\,\frac{ \ln(\xi/\xi_0)}{\ln(\sqrt{Dt}/\xi_0)}\,,
\end{equation*}
and estimate the closest particle distance $\ell$ from the extreme statistics
criterion
\begin{equation*}
 \int_{\xi_0}^\ell d\xi\int_0^\ell d\eta \int_0^{2\pi} d\phi\,\, 
 \frac{\xi+\eta}{4}\,\rho\,\frac{ \ln(\xi/\xi_0)}{\ln(\sqrt{Dt}/\xi_0)}\sim 1\,,
\end{equation*}
where $\tfrac{1}{4}(\xi+\eta) d\xi\, d\eta\, d\phi$ is the volume element in
parabolic coordinates.  Computing the integral we find
\begin{equation*}
\frac{\rho\ell^3\,\ln(\ell/\xi_0)}{\ln\tau} \sim 1\,, \qquad 
\tau = \frac{Dt}{\xi_0^2}
\end{equation*}
Hence the leading asymptotic is
\begin{equation}
\ell \sim \left[\frac{\rho^{-1}\,\ln\tau}{\ln(\rho^{-1}\xi_0^{-3}\ln\tau)}\right]^{1/3}\,.
\end{equation}

When $d\geq 4$, the density remains finite and essentially uniform, as
particle depletion occurs only over distances of the order of $\xi_0$ (see
Eq.~\eqref{S:finite}).  Thus if the initial density is low, $\rho\xi_0^d\ll
1$, the distance between the closest particle and the apex remains of the
order of $\rho^{-1/d}$, in agreement with the naive prediction of
Eq.~\eqref{dist:naive}.  Summarizing, 
\begin{equation}
\ell \sim 
\begin{cases}
(Dt)^{1/10}\, \rho^{-2/5}         &d=2\,,\\
(\ln\tau)^{1/3}\,[\rho\ln(\rho^{-1}\xi_0^{-3}\ln\tau)]^{-1/3}    &d=3\,,\\
\rho^{-1/d}                      &d\geq 4\,.
\end{cases}
\end{equation}

As indicated by Fig.~\ref{l}, there are two natural measures of closest
distance, both of which are needed to characterize the outer envelope of the
spatial distribution of surviving particles.  Very far away from the apex of
the absorbing paraboloid, namely, for distances much greater than $\ell$, the
paraboloid is locally planar.  In this limit, the distance to the closest
particle, defined as $b$ in Fig.~\ref{l}, is given by the third line of
Eq.~\eqref{l-general}.  Thus for the specific case of the parabola ($d=2$),
we have
\begin{equation*}
  \ell   \sim (Dt/\rho^4)^{1/10}\,,  \qquad\qquad
  b \sim (Dt/\rho^2)^{1/6}\,.
\end{equation*}
These two distances help characterize the spatial distribution of surviving
particles without requiring the full solution of the diffusion equation.

\section{Discussion}

The first-passage properties of a diffusing particle inside an absorbing
paraboloid represents a natural and phenomenologically rich extension to the
first-passage properties inside an absorbing cone.  For the conical system,
the survival probability $S(t)$ generally decays as a non-universal power law
in time and a basic question, now largely settled~\cite{bk1}, has been to
compute the exponent of this power law as a function of the cone angle and
spatial dimension.  An intuitive way to understand why there should be a
non-universal power-law decay is to decompose the diffusive motion
longitudinally and transversely.  Longitudinally, a diffusing particle
wanders a distance $y\sim \sqrt{t}$ in a time $t$.  At this elevation, the
transverse distance to the boundaries is $x= y\cot\alpha$ which is also
growing as $\sqrt{t}$.  This physical picture leads to a survival probability
that has a non-universal power-law time dependence~\cite{kr}.

For the paraboloid defined by $y=aR^p$, where $R$ is the transverse radius
(Eq.~\eqref{p:paraboloid}), the same decomposition of the motion into
longitudinal and transverse components shows that the range of the transverse
interval now grows more slowly than $\sqrt{t}$.  The feature leads to a more
rapid than power-law decay of the survival probability as a function of time.
Making use of this decomposition, we constructed an extreme statistics
argument that predicted $\ln S\sim -t^{\frac{p-1}{p+1}}$.  The exponent value
agrees with rigorous bounds and calculations~\cite{bds,ls}; we explained how
to deduce this result with little effort.

We also studied the mean time until trapping inside the paraboloid $y=aR^2$.
This mean time, as well as all higher moments, are finite, in distinction to
the corresponding behavior inside an absorbing cone.  As is naturally
expected, the trapping time for a diffusing particle that starts at a given
point inside the paraboloid increases as the square of the distance to the
closest point on the paraboloidal surface.  There are two natural extensions
of this computation that have not yet been done: calculating the mean time
until trapping inside non-quadratic paraboloids and determining the full
distribution of trapping times.  To the best of our knowledge this
distribution has not been computed even for circular cones.

Finally, we investigated the time dependence of the distance $\ell$ between
the apex of an absorbing cone and the closest particle for an initially
uniform particle distribution.  We found the $\ell$ increases as a power-law
in time, with exponent that depends on the wedge opening angle.  The increase
in this minimum distance becomes extremely slow as the wedge opens up into an
excluded half line in two dimensions.  In contrast, in three dimensions, the
closest distance to an excluded half line increases only logarithmically with
time.  These latter behaviors were exploited to determine the closest
distance between the apex of an absorbing paraboloid and surviving diffusing
particles.

\begin{acknowledgements}
  We thank Eli Ben-Naim for collaboration on related subjects that helped to
  inspire this project.  This work has been supported by NSF grant CCF-0829541
  (PLK) and NSF grant DMR0906504 (SR).
\end{acknowledgements}


\begin{thebibliography}{99}
\bibitem{location} The precise location of the starting point does not affect
  the most interesting asymptotic behaviors.
\end{thebibliography}
\end{document}